\documentclass[conference]{IEEEtran}
\IEEEoverridecommandlockouts
\usepackage{cite}
\usepackage{amsmath,amssymb,amsfonts}
\usepackage{algorithmic}
\usepackage{graphicx}
\usepackage{textcomp}
\usepackage{xcolor}
\def\BibTeX{{\rm B\kern-.05em{\sc i\kern-.025em b}\kern-.08em
    T\kern-.1667em\lower.7ex\hbox{E}\kern-.125emX}}
\begin{document}

\makeatletter
\newcommand{\linebreakand}{%
  \end{@IEEEauthorhalign}
  \hfill\mbox{}\par
  \mbox{}\hfill\begin{@IEEEauthorhalign}
}
\makeatother

\title{Combining Node-RED and Openwhisk  for Pattern-based Development and Execution of Complex FaaS Workflows
}

\author{\IEEEauthorblockN{George Kousiouris}
\IEEEauthorblockA{\textit{Dept. of Informatics} \\
\textit{Harokopio University}\\
Athens, Greece \\
gkousiou@hua.gr}\\
\and
\IEEEauthorblockN{Szymon Ambroziak}
\IEEEauthorblockA{\textit{GFT} \\
Lodz, Poland \\
Szymon.Ambroziak@gft.com}\\ 
\and
\IEEEauthorblockN{Domenico Costantino}
\IEEEauthorblockA{\textit{Pointnext Advisory and Professional Services} \\
{Hewlett-Packard Italiana Srl} \\
Cernusco s/N, Milan, Italy \\
costantino@hpe.com}
\and
\IEEEauthorblockN{Stylianos Tsarsitalidis}
\IEEEauthorblockA{\textit{Dept. of Informatics} \\
\textit{Harokopio University}\\
Athens, Greece \\
styltsars@gmail.com}\\ 
\linebreakand %
\IEEEauthorblockN{Evangelos Boutas}
\IEEEauthorblockA{\textit{Dept. of Informatics} \\
\textit{Harokopio University}\\
Athens, Greece \\
it21662@hua.gr}\\ 
\and
\IEEEauthorblockN{Alessandro Mamelli}
\IEEEauthorblockA{\textit{Pointnext Advisory and Professional Services} \\
{Hewlett-Packard Italiana Srl} \\
Cernusco s/N, Milan, Italy \\
alessandro.mamelli@hpe.com} \\ 
\and
\IEEEauthorblockN{Teta Stamati}
\IEEEauthorblockA{\textit{Dept. of Informatics} \\
\textit{Harokopio University}\\
Athens, Greece \\
teta@hua.gr}
}
\maketitle

\begin{abstract}

Modern cloud computing advances have been pressing application modernization in the last 15 years, stressing the need for application redesign towards the use of more distributed and ephemeral resources. From the initial IaaS and PaaS approaches, to  microservices and now to the serverless model (and especially the Function as a Service approach), new challenges arise constantly for application developers. This paper presents a design and development environment that aims to ease application evolution and migration to the new FaaS model, based on the widely used Node-RED open source tool. The goal of the environment is to enable a more user friendly and abstract function and workflow creation for complex FaaS applications. To this end, it bypasses workflow description and function reuse limitations of the current FaaS platforms, by providing an extendable, pattern-enriched palette of ready-made, reusable functionality that can be combined in arbitrary ways.  The environment embeds seamless DevOps processes  for generating the deployable artefacts  (i.e. functions and images) of the FaaS platform (Openwhisk). Annotation mechanisms are also available for the developer to dictate diverse execution options or management guidelines towards the deployment and operation stacks. The evaluation is based on case studies of indicative scenarios, including creating, registering and executing functions and flows based on the Node-RED runtime, embedding of existing legacy code in a FaaS environment, parallelizing a workload, collecting data at the edge and creating function orchestrators to accompany the application. For the latter, a detailed argumentation is provided as to why this process should not be constrained by the "double billing" principle of FaaS.       

\end{abstract}


\section{Introduction}

In recent years, infrastructures and services have been characterized by the continuous advancements in the area of cloud computing, starting from the use of ephemeral and elastic virtual machines and reaching the level of lightweight containerization approaches for application packaging and deployment. In order to adapt and exploit  this new way of resource provisioning, application developers had to go through an immense adaptation process, having to tackle issues like state handling, ephemeral nature of the resources, need for more distributed and elastic application behaviour, as well as avoid common design pitfalls in the microservice domain \cite{neri2020design}. 

Architectural approaches such as  microservice-based design principles \cite{nadareishvili2016microservice} or cloud native application design considerations\cite{Design} have aided as a guide in the process but have not alleviated the developer from the main effort needed to break down their applications as well as handle the aforementioned issues. 
What is more, with the advent of further computing models like serverless computing and specifically Function as a Service\cite{lynn2017preliminary}, the developers are yet again facing challenges to migrate the core of their applications into more fine-grained, function oriented chunks. 

Indicative challenges\cite{abad2021serverless}  in the new computing model include extension of the application domain through a suitable trade-off between  expressivity (of the application graph) and simplicity, maintainable composition models for  serverless workflows, usage of patterns for serverless applications, inclusion and support for the legacy part of serverless applications, versatile development processes, supported by relevant development tools and CI/CD processes. 

 Other approaches\cite{kousiouris2021functionalities}) also suggest the application development continuum approach for combining functions and services in one environment, using annotations to propagate dictations and managerial approaches directly from the developer to the underlying management layers, abstractions  and visual development  tools  for  building non-trivial  FaaS  applications, while combining and adding legacy parts to short-lived functions without the need for extensive application refactoring. The addition of programming  patterns  offered as reusable components  may  significantly  aid application   adaptation and functionality. 

The purpose of this paper is to present the PHYSICS cloud Design Environment for FaaS, a framework for enabling easier application workflow creation and adaptation  to the FaaS model.  The PHYSICS environment encapsulates the widely used (in the IoT domain) Node-RED\footnote{https://nodered.org/} web based function framework for event driven applications, coupled with a back-end system that undertakes the preparation of the provided code and functionality for registration and deployment to a target FaaS platform (based on Openwhisk). 

The goal of the environment is to provide the following contributions:

- user-friendly visual way of creating functions and linking them together in workflows, exploiting a palette of existing functionality from the Node-RED environment

- provide a set of implemented functionalities in the form of patterns (subflows), that may aid the developer in the adaptation to the FaaS paradigm, used directly in a drag and drop manner in the workflow

- support a wide variety of execution modes through a modular DevOps process, including the ability to execute the created flow as a native sequence on the target platform, as one Openwhisk function (based on the Node-RED runtime), including an entire flow of functions, as an orchestrator function (orchestrating other deployed functions), or as a service 

- include diverse annotations as guidelines to other management components down the stack, enabling extended options such as function placement, preferences in scheduling, inclusion of an accompanying service component etc.

Through the aforementioned features, application creation, adaptation and migration to the FaaS platform can be significantly enhanced, limiting the learning curve and development time. The paper proceeds as follows. Section II presents related work, both in terms of open source and commercial FaaS platform capabilities from a workflow creation point of view, comparing them to the baseline Node-RED capabilities. Section 3 presents the main architecture and building blocks of the PHYSICS Design Environment, as well as the rationale of the execution modes. Section 4 presents a set of example case studies and provided capabilities of the environment while  Section 5 concludes the paper.

\section{State of the Art on FaaS platform environments}

According to \cite{leitner2019mixed}, a major drawback of current Function as a Service platforms is the availability of tools  related  to deployment and function reuse. Function composition however can be used in order to provide more complex groups of functions through the interaction of simpler ones, provided that there are according orchestration and grouping capabilities\cite{amato2017exploiting}. The notion of patterns can in this case be useful in order to group appropriate functions that intend to solve a specific problem (e.g. AI training and optimization\cite{giampa2020mip}).   

From the main open source platforms, 
Openwhisk\footnote{https://openwhisk.apache.org/} is the only platform that  has a native functionality in place (sequence operator) at the runtime level\cite{baldini2016cloud}, although it only supports simple sequences of function chains. A further extension involves the IBM Composer\cite{Composer} which implements a set of orchestration primitives, although in a javascript library based form\cite{barcelona2019faas}. The functionality is also ported to IBM Cloud Functions, the commercial cloud solution from IBM, one of the main contributors of the Openwhisk project.

OpenFaaS has an external plugin (FaaS-flow\cite{FaaSflow}) for declaring sequences of functions in a textual,code-like manner (including more complex workflow primitives). The orchestrating logic is also executed as a function. Kubeflow\cite{bisong2019kubeflow} has the pipeline definition language as well as an editor extension (Elyra) through which pipelines can be visually defined. However the concept of workflow in this case is that of a static sequence of operations (with information passing from one step to another via intermediate cloud object storage files). Therefore it lacks the dynamicity and the abilities of an actual runtime environment correlating the passing of arguments between functions and writing arbitrary orchestration code. 

From the main cloud vendors, AWS Step functions includes a visual programming style, as well as a number of further operators (including state management), however it is directly tied to the AWS services and thus is an option that increases vendor lock-in. Google Cloud Functions supports the creation of workflows through relevant yaml files and syntax\cite{Google}, assuming that the functions have already been deployed. In Fig. \ref{fig:gcfsample}, the comparison between a GCF based syntax (left) and an equivalent Node-RED flow (right) implementing the same functionality indicates the differences in the usability of the two forms, even for simple flows. In the Node-RED flow, only the ready-made client nodes\cite{OWnode} for the FaaS platform (in this case Openwhisk) are needed (with the name of the function to use) plus a small custom function for adapting  message fields. Going to even more complex workflows, including a large number of functions and diverse connections between them, becomes tiresome and error prone for the developer in the YAML format, as well as in the other text/code based ones such as IBM Composer or FaaS-flow. The comparison performed in \cite{barcelona2019faas} indicates drawbacks  of workflow management abilities among cloud providers in the case of fork-join primitives, with the exception of IBM Composer.

\begin{figure*}[htbp]
\includegraphics[width=\textwidth]{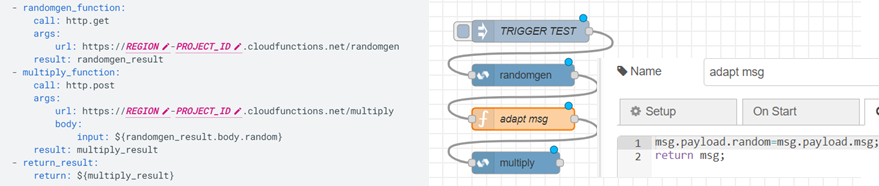}
\caption{Example code snippet for workflow definition in Google Cloud Functions\cite{Google} compared to equivalent Node-RED defined sequence }
\label{fig:gcfsample}
\end{figure*}
\textit{Baseline Node-RED environment description and comparison}

Apache Airflow is a workflow design and management tool. It allows to plan, schedule and automate the flow of data through nodes. Graphs of Directed Acyclic Graphs (DAGs) type represent the direction of the data, while the output of one node (task) is usually the input of the next node. Airflow has complex operators, however it can not be used as a generic function editor. One very interesting work is TriggerFlow\cite{arjona2021triggerflow}. In this case different workflow primitives are offered, as well as eventing mechanisms in order to regulate the execution. The main difference of our work is that in our case the  environment can be used for both function as well as workflow creation. Furthermore, due to the usage of Node-RED, we are able to import ready-made functionality in the form of reusable subflows and nodes, either at the workflow structure or at the core functionality level. Another difference is the visual support for the workflow creation. 

What can be observed from the investigation of the related work is that a number of solutions exist, but primarily for the definition of workflows either without a proper runtime mechanism, or through code-level libraries/yaml files difficult to write for flows containing many functions and complex connections. They have very little support for function re-use (especially function groups reuse) and no support for annotations that can somehow be propagated to lower levels of management.  There is no single environment that can help the developer visually design, test and deploy the functions while also acting as an orchestrator during runtime. 

From that aspect, Node-RED portrays a number of significant advantages in terms of the aforementioned features if one wants to use it as a design and development environment for FaaS:

- a visual server environment for wiring and deploying together functions into complex and arbitrary workflow structures, without the aforementioned limitations in terms of types of wirings or implemented orchestration patterns. The programming style of the environment follows a functional, event driven programming approach, fitting to the baseline FaaS model.

- a more augmented runtime environment that can be used as the basis of execution,  aiding in creating FaaS functions (or actions) that internally consist of multiple Node-RED functions. This enhances the development process (due to the runtime abilities for message tracking and manipulation) as well as enables local flow testing directly in the Node-RED server editor, skipping the costly and time consuming actual FaaS deployment. Furthermore, the inclusion of more than one Node-RED functions in a FaaS function output would result in less needs for containers during execution, therefore less back-end contention. 

- abundance of ready-made nodes\cite{Noderedrepo}, especially from the IoT domain but also for general systems, exploiting the generic npm repository of nodejs, one of the largest open source repositories

- ability to group functions as subflows, aiding in code reusability, sharing and function management, workflow simplification and abstraction. 

-ability to treat the workflow definition as a meta-specification layer. Given that Node-RED has its own simple workflow definition schema, this can act as a meta-specification from which translations to different provider syntaxes can be performed in order to mitigate vendor lock-in.

Node-RED typically runs as a server, so the main question is: can it be used initially as an editor and testing environment for creating functions and workflows that are afterwards deployed on a typical FaaS platform? For this purpose, further backend services and functionalities need to be offered. Furthermore, can it also be used for orchestrating functions and in what way? And what happens in this case with common issues like double billing in the serverless trilemma\cite{baldini2017serverless}, i.e. the principle in which no function should wait (and get billed) while waiting for another function to complete.

\section{PHYSICS Design Environment Architecture}

\subsection{Design Environment Overview and Main Architectural Blocks}

The overview of the PHYSICS Design and Development Environment appears in Fig. \ref{fig:generic}. The main editing environment is an embedded container of a Node-RED server, with an enriched palette of nodes (including the built-in ones, additions from the Node-RED community repository as well as extensions provided by the PHYSICS environment). The PHYSICS editor extensions include either ready-made subflows that are built for a specific purpose (see section III.B for details) or semantic annotation nodes (see section III.C for details). All these elements can be drag and dropped directly in an arbitrary workflow in order to augment its functionality.

\begin{figure*}[htbp]
\includegraphics[width=\textwidth]{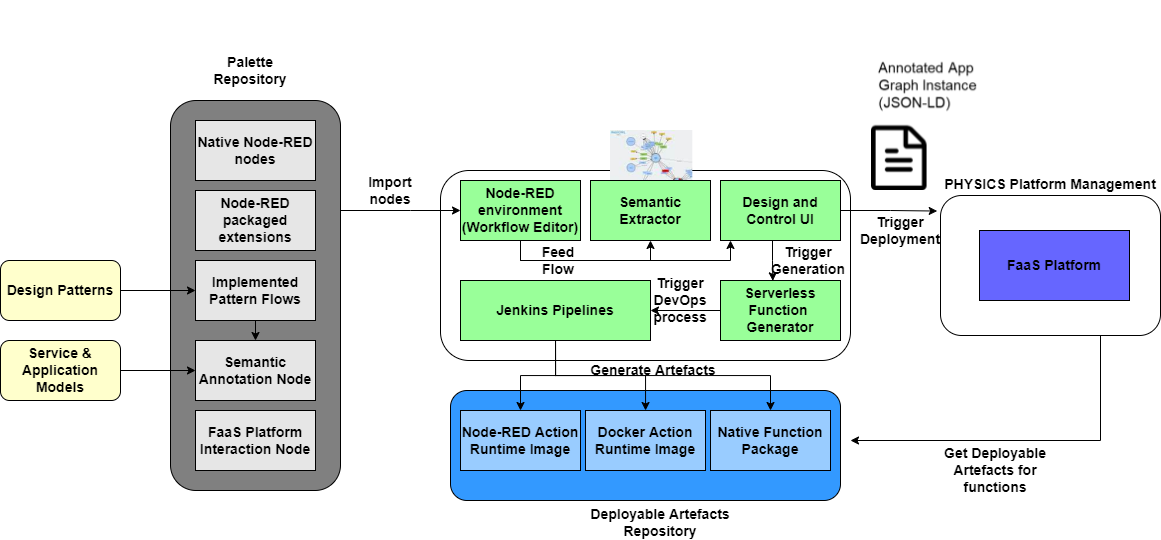}
\caption{Overview of the PHYSICS Design and Development Environment}
\label{fig:generic}
\end{figure*}

Once the developer finishes the development of the flows, they can move to the Design and Control UI, in which they select which flows to prepare for instantiation. From then on, the process is orchestrated by this component, which contacts the Serverless Function Generator for extracting the code from the Node-RED environment and calls a relevant DevOps process implemented as a Jenkins pipeline. Different pipelines are supported based on the mode of desired execution (native function or Node-RED runtime function) in order to adapt to the needed steps in each case. 

The flow passes also through the Semantic Extractor component, which extracts the declared annotations in the selected flows and maps them to triples based on ontological concepts. The triples are stored for later use, embedded  as JSON-LD declarations in the overall application graph (the JSON specification of the typical Node-RED flow). 

Upon finalization of this process, the annotated app graph is forwarded to the PHYSICS platform management layer, that includes functionality in order to process the graph and register the according functions and sequences to the FaaS platform. It also maintains the relevant annotations (e.g. mapped to Kubernetes keywords) in order to dictate the expressed developer needs (in terms of deployment, function management, affinity, sizing etc.). Following, details on the different execution patterns are presented.

\subsubsection{Execution of Node-RED functions as native Openwhisk functions}

The functions created within the Node-RED environment can be executed directly as according functions in one of the native runtimes of Openwhisk (nodejs,python, java etc.). In order to achieve that, one needs to go through a packaging stage from Node-RED functions to native Openwhisk functions. For example, for the javascript functions the code is extracted from the Node-RED javascript function node in a single JavaScript function named “main” in main.js file of a simple npm project. In addition libraries used need to be retrieved and installed as dependencies of the npm project. After that the whole project is packaged as a zip file, which can be deployed directly to OpenWhisk. In order to support native sequences of Openwhisk, a simple transformation from the JSON specification of the Node-RED flow (indicating wires between nodes) needs to be made to the sequence specification of Openwhisk. At this stage, javascript and python native functions are supported. 

\subsubsection{Node-RED as function execution runtime process}

The major ability however that interests most is the ability to write any arbitrary workflow in Node-RED, expoiting any type of node, that can be then executed either as a service (typical Node-RED execution) or as a function. This process is split into two parts. Initially an in-flow support is provided, in the form of a skeleton flow. For example, Openwhisk assumes that any registered function artefact exposes two endpoints (a POST /init method and a POST /run method) used to initialize (if needed) and then execute the function logic. Inside this flow, any node-RED packaged node as well as npm-based library can be exploited, thus leading to the inclusion of a code base with extensive capabilities. In this mode, the developer can wire nodes in whatever manner, since their execution is performed within the Node-RED runtime of the Openwhisk action, not limited by any workflow specification limitation of the FaaS platform. 

An extra benefit of this case is that it solves the issue of splitting the function logic packaging in many different separate functions, which may lead to difficulty in managing hundreds of functions in the context of one application, as identified in \cite{abad2021serverless}. Furthermore, grouping many lightweight, trivial functions in one actionable function image, based on the Node-RED runtime, results in much fewer container overheads due to the fact that no separate container needs to be raised for each individual and potentially small function. 

The second needed step is the creation of the deployable artefact. In this case, the environment pushes all the changes made by the developer in the main Node-RED editor container as well as gets the content of the selected flows from the Node-RED API. Then the function docker image is built, exploiting the same base image of the Node-RED development environment, by replacing the default flows with the target flow and utilizing any environment settings files needed (e.g. credentials, settings and keys). After that, the image is pushed to a Docker registry, from which it can be directly deployed to Openwhisk. The Jenkins pipelines can also be enriched with function registration to a test FaaS platform. 

\subsection{Pattern-based Flow creation}
A pattern is defined as \textit{“a proven series of activities which are supposed to overcome a recurring problem in a certain context, particular objective, and specific initial condition”}\cite{wahyudi2018process}. Patterns have been a very useful tool for dictating design principles as well as driving abstract implementations for a specific domain \cite{jamshidi2017pattern}. A very thorough list of patterns for the cloud domain can be found in \cite{Design}, in which details for each pattern include its design, its parameters, when and how it should be used, benefits and drawbacks etc. The same applies for FaaS-specific patterns, a list of which can be found in \cite{taibi2020patterns}. In general a pattern may be driven either from research goals, the peculiarities of a specific domain or the challenges of a specific use case/application.    

In the context of this work, the concept of a pattern fits very well with the generic ability of Node-RED to group entire workflows in subflow nodes and add properties and configurations on top of them. This functionality then is included in the environment palette as a single node, hiding all underlying complexity. The developer can directly drag and drop that node in their flow and use it directly, a feature that significantly speeds up development. 

As an example, we present the created split (fork)-join pattern to parallelize embarrassingly parallel computations based on the Single Instruction Multiple Data (SIMD) pattern (Fig.   \ref{fig:splitjoin} a). These computations are in many cases based on typical parallel technologies like MPI, which however are more difficult to develop and support. In a function programming style, the parallelization may be performed by splitting the initial message, that contains an array of rows upon which the same computation will be applied on each row. Each split message then triggers a respective function execution on a FaaS platform, while a Join node waits for all the respective partial messages created from the original message. 
\begin{figure*}[htbp]
\begin{center}
\includegraphics[scale=0.25]{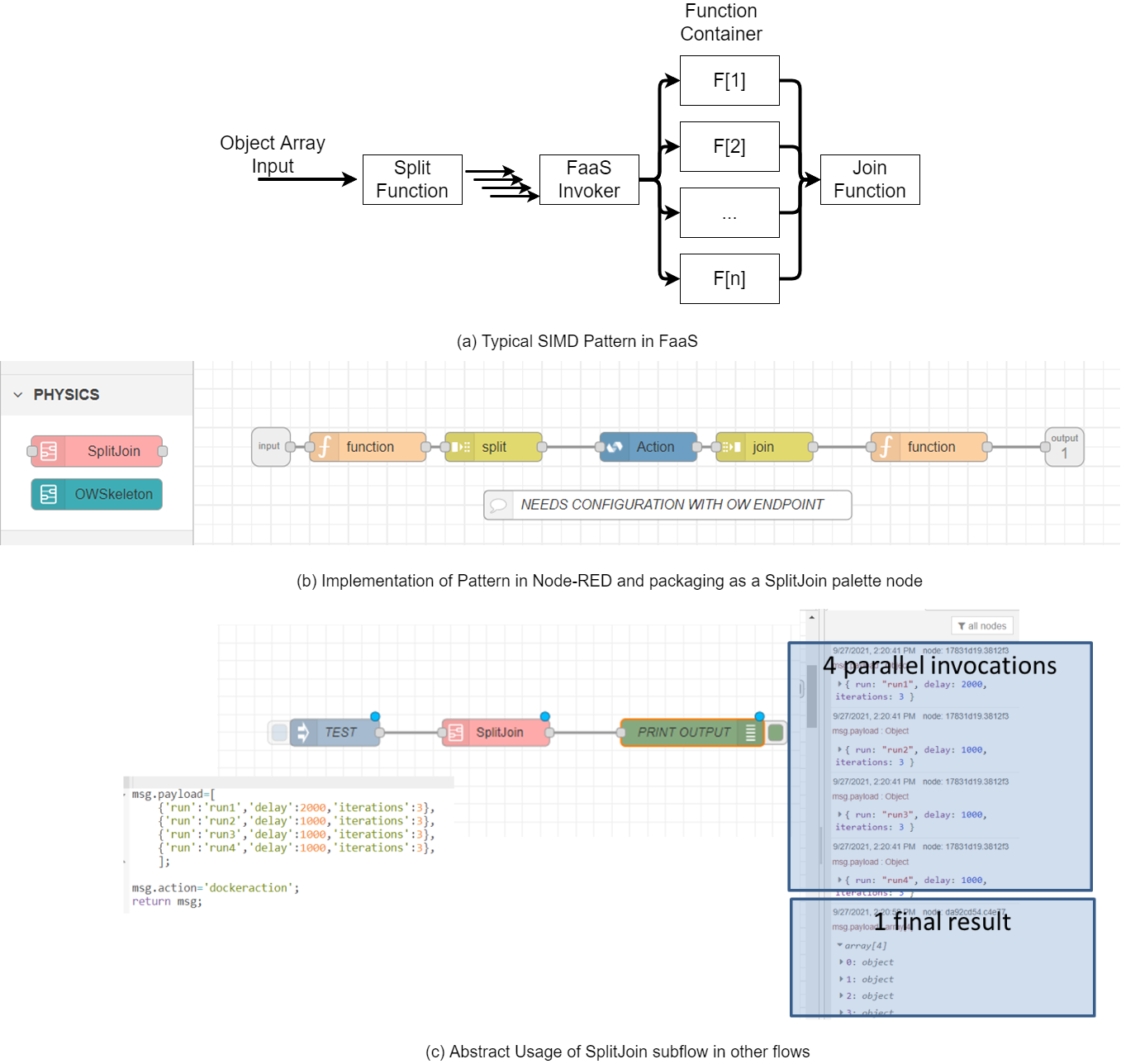}
\caption{Example Split(Fork)-Join Pattern}
\label{fig:splitjoin}
\end{center}
\end{figure*}

The implementation in Node-RED appears in Fig. \ref{fig:splitjoin}b. Grouping at the output is performed based on the unique original message id that characterizes all partial messages as well as a msg.parts field that indicates the position in the initial sequence of each partial message. The Split and Join nodes are built-in Node-RED nodes. The Action node is a Node-RED packaged node available on npm\cite{OWnode} to interface with an existing Openwhisk platform, while the "function" nodes include custom code used to adapt incoming and outgoing message fields (e.g. enforce a convention that the array to be split is included in the msg.payload field). 

The overall functionality is included in a subflow (SplitJoin node in left-side palette), and can be used directly in an example flow (Fig. \ref{fig:splitjoin}c) with minimum effort, indicating only the Openwhisk endpoint and the split size (how many rows in the initial array will be included in each FaaS function input). The latter can be configurable in the Split node and is useful for not creating too many function invocations that would result in large container numbers. By using a larger split size, i.e. feed many input rows in one function execution, this aspect can be optimized.    

It is not necessary for a pattern to be only implemented in a function oriented style, in many cases a service implementation can also be used as a pattern in serverless environments, as the example batch request aggregation pattern in \cite{Kousiouris2021self}.

\subsection{Semantic annotations inclusion}

Semantic annotations can give a wide number of possibilities to the developer in order to affect various aspects of a flow or function execution. As an example, selection between the execution modes (Node-RED flow function, native function or service) can be performed at the flow level, while other considerations may include the deployment target (e.g. function or flow A needs to be deployed on Edge B), direct inclusion of existing deployable artefacts, priority scheduling if supported by the underlying platform, affinity rules, QoS constraints, needed H/W  etc.

In the PHYSICS environment, there are currently two ways to import annotations. At the function level, in-code annotations wrapped around a specific syntax (//@key=value), following the example of  annotations in the Dependency Aware FaaSifier implementation\cite{ristov2020daf}. At the flow level, a special set of nodes (semantic annotators) has been created as subflows and included in the palette. In this case the required fields can be included easily as subflow properties and environment variables. The instantiation of such a subflow remains in the JSON specification of the Node-RED flow (along with the set values) and can be processed by the Semantic Extractor component mentioned in Section III.A. An example of such a node appears in Fig. \ref{fig:imageimport}, in which the developer may declare an external docker image as an actionable Openwhisk function, for inclusion in the registration process. Examples of in-code usage may include the use of the //@locality=edgeA annotation in order to indicate that a function should be deployed on that particular location.

\begin{figure}[htbp]
\begin{center}
\includegraphics[scale=0.4]{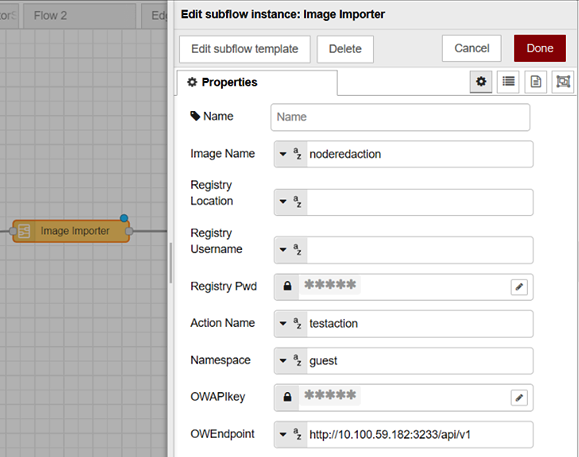}
\caption{Example of Image Importer annotation node}
\label{fig:imageimport}
\end{center}
\end{figure}

\subsection{Orchestrator Flows for complex function workflows}

Due to the limitations of the reviewed environments and the inherent ability of Node-RED to act as an orchestrator, one key feature is to use a Node-RED flow in order to orchestrate complex function wiring and workflow primitives, exploiting the Node-RED Action runtime presented previously and the execution of a flow as a function. This has the extra benefit that the specific orchestration definition, based on the Node-RED workflow meta-specification, can be executed on multiple providers with limited changes (e.g. only on the skeleton flow for each platform), by adapting to the underlying workflow specifications used by each provider. This directly leads to a decreased vendor lock-in for the case of FaaS.

The created orchestration flow can be deployed either as a function or as a service, exploiting the relevant annotation (example in section IV.D). However in the service mode one is constrained  by the scalability of a single Node-RED environment used to orchestrate many executions. Furthermore, they get billed for the constantly running orchestrator service. These two arguments are the most commonly used for moving a functionality to a serverless paradigm in any case. The two potential orchestration ways appear in Fig. \ref{fig:orch_mode}

\begin{figure}[htbp]
\begin{center}
\includegraphics[scale=0.25]{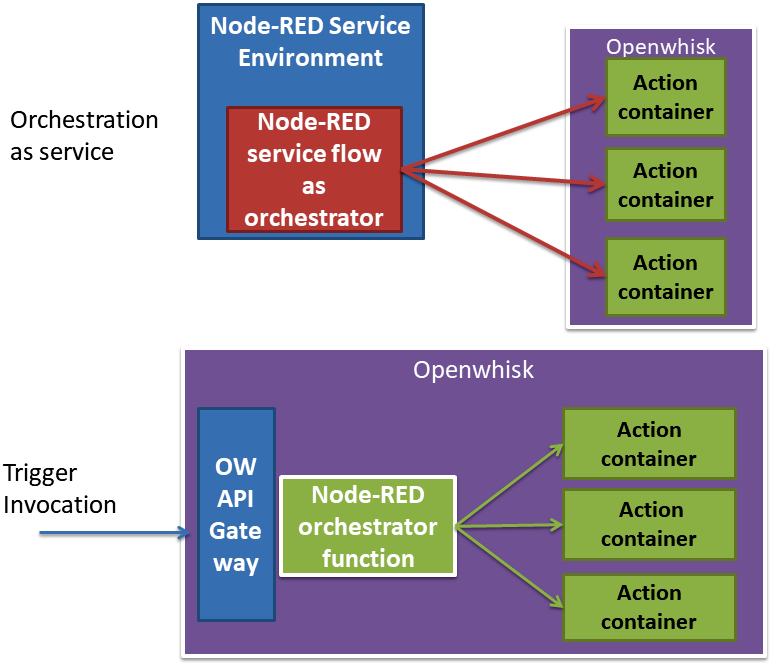}
\caption{Service and Function Orchestration Modes}
\label{fig:orch_mode}
\end{center}
\end{figure}

Despite the aforementioned advantages, usage of a Node-RED flow acting as an orchestrator and executed as a function would result in a double billing\cite{baldini2017serverless} issue, since the orchestrator function would need to wait for the orchestrated functions to finish. However there are various arguments for following this approach, presented in the next section.

\subsubsection{Argumentation against the double billing principle in Orchestrator Functions}

The main arguments against considering orchestration as part of the double billing principle are the following:

- whether the orchestrating function would be billable or not is primarily a business model decision and should be separated from the technical ability. I.e., the provider might choose to offer such a functionality for free or with a different cost model than function execution time. If the respective provider wants to gain a competitive advantage and give the ability to their customers to easily create and deploy arbitrary workflows, they could follow this approach. In a similar manner, many of other types of services (like usage of dashboards) are not separately billed by the providers, even though there are available REST APIs for every possible action performed through the dashboard. The aforementioned services are offered primarily as a more user friendly mean to create and manage cloud resources, thus leading to increased uptake of the domain. Similarly, in the serverless domain specifically, there is no charge for e.g. the gateway service that is running and listens for events that trigger functions.   

- usage of such an approach would alleviate from the need for a scalable orchestrator, a daunting issue on its own, since each separate orchestrated flow would be in its separate function execution. This setup  is by default scalable. Furthermore, the existence of an actual orchestrating runtime (Node-RED runtime) would mean that any workflow primitive could be applied based on custom logic and appropriate message handling.

- according to \cite{eismann2021state}, 82\% of serverless applications use 5 or fewer functions and only 31\% of them include workflows. Even these are mostly simple structured, small, and short-lived. Having the ability to create more complex application workflows would be in favour of the providers in the long run, since more complexity in the workflows would directly incur higher number of included functions and according invocations. 

- if we consider that a function should not wait for an operation (i.e. blocking call) since in that time it is billed without being useful, why do we accept blocking calls in the most typical serverless use cases, e.g. the retrieval of a data object from an object storage prior to feeding it to an AI image detection function, and not go for alternative near-data processing models\cite{montagnuolo2016supporting}. 

In a nutshell, we consider the fact that the double billing principle can aid when writing single functions. It leads to a more asynchronous style of programming, however it should not be applied at the function workflow orchestration level, given the constrained current abilities of FaaS platforms in this feature, as detailed in Section II.

\section{Example Case Studies for Approach Validation}

In order to test and experiment with the environment, a number of case studies have been implemented and described in the following paragraphs.

\subsection{Example A- Simple Openwhisk Skeleton Wrapper for Node-RED and investigation of Node-RED runtime perfomance}

One of the execution modes as mentioned in Section III includes the ability to define a Node-RED flow that will be wrapped and executed as a function in Openwhisk, either in a native nodejs or a Node-RED runtime. The Openwhisk specification indicates that any docker image can be included as a function provided that it has a REST interface with two endpoints (a POST /init and a POST /run). The /init method is used to initialize the environment of the function (if needed) while the /run is used for the actual execution of the function and for passing any input arguments.

A relevant subflow has been created (OWSkeleton) that includes the two methods if one wants to follow the Node-RED function execution mode (Fig. \ref{fig:helloflow}). One question is how different in terms of performance the execution of a Node-RED function is, compared to the execution based on the typical, built-in nodeJS runtime. To this end the OWskeleton flow was used with a hello world function. A similar hello world implementation was created and registered for the typical lightweight nodejs runtime of Openwhisk. The Openwhisk setup included a single node with 8 cores and 16 GB of RAM, given that we were primarily interested in the baseline times of one function execution.

\begin{figure}[htbp]
\includegraphics[width=\columnwidth]{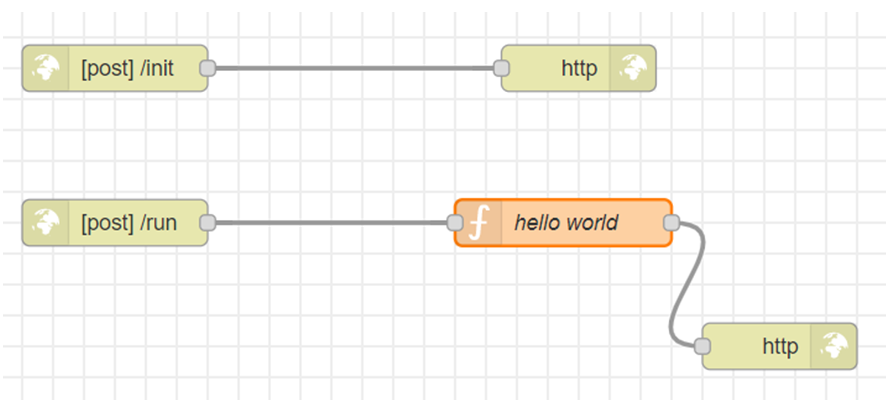}
\caption{Hello World Flow based on OW skeleton template}
\label{fig:helloflow}
\end{figure}\textbf{}
Afterwards, 100 test runs were conducted on each target function, to measure the cold start times (need to spawn a new container) and the hot ones (reuse of an existing container). The results appear in Fig. \ref{fig:combined}. As anticipated, the cold start time for the Node-RED runtime is considerably higher (with an average of 7.6 seconds compared to 2.5 seconds), given that it is based on a larger container with more dependencies, as well as the fact that the Node-RED environment has a more heavyweight start-up time during function execution. In the hot execution case, the two execution modes are very similar, with Node-RED having a slight advantage (227 against 252 milliseconds) as well as smaller deviation. Addressing the larger cold start times could also be performed by utilizing the Function Warmer pattern\cite{taibi2020patterns}. However with the use of the Node-RED function, one gains significant abstraction in development, as well as the benefit of the ready-made nodes.

\begin{figure}[htbp]
\includegraphics[width=\columnwidth]{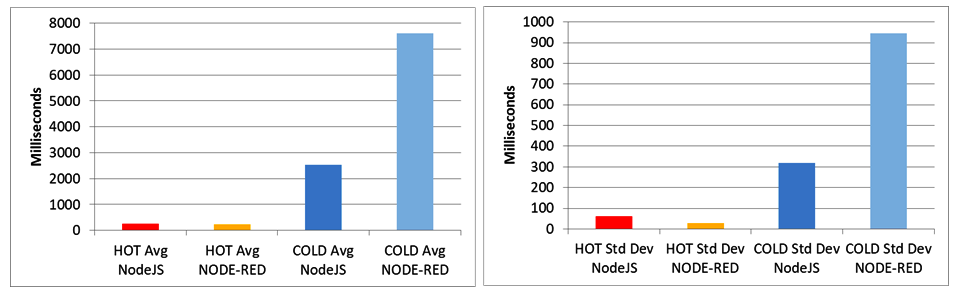}
\caption{Hello World Function Average and Std Dev Times for NodeJS and Node-RED runtimes}
\label{fig:combined}
\end{figure}

\subsection{Example B- Inclusion of a Docker image based legacy component }

In order to test the porting of an existing legacy component, we selected the AI service that was used in the context of \cite{Kousiouris2021self}. This consists of a REST endpoint that is responsible for obtaining requests to perform model inference through the launching of containerized models embedded in an Octave environment container. 

For adapting the respective implementation, the baseline image, consisting primarily of the GNU Octave environment and relevant scripts for interacting with the model creation and querying process, was enriched with the Node-RED runtime. A relevant flow was also created (Fig. \ref{fig:octave}) , utilizing the Openwhisk Skeleton as well as internal Node-RED functions in order to adapt the arguments passed to the Octave scripts. While in the previous version these arguments were passed through Docker commands as environment variables, in this case the adaptation included passing them directly as command line arguments of the scripts, after obtaining them through the Openwhisk argument interface (body of the POST /run call).
\begin{figure}[htbp]
\includegraphics[width=\columnwidth]{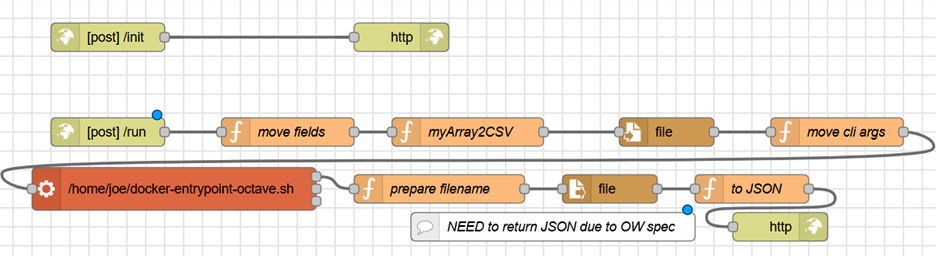}
\caption{Adaptation Flow of AI model inference service from \cite{Kousiouris2021self}}
\label{fig:octave}
\end{figure}

The difference between the time delays of the previous versus the serverless implementation appears in Fig. \ref{fig:octave_measurement}, for 150 runs in a frequence of 1 request per second in order to match the measurements from \cite{Kousiouris2021self}. It is indicative that due to the reuse of the already spawned containers, a typical feature of Openwhisk, a 20x benefit in terms of execution time is gained for the response time of the FaaS service, when compared to the original service version that spawned a new container for each request. The adaptation process was also simple, needing less than 2 working days to be performed. 
\begin{figure}[htbp]
\begin{center}
\includegraphics[scale=0.25]{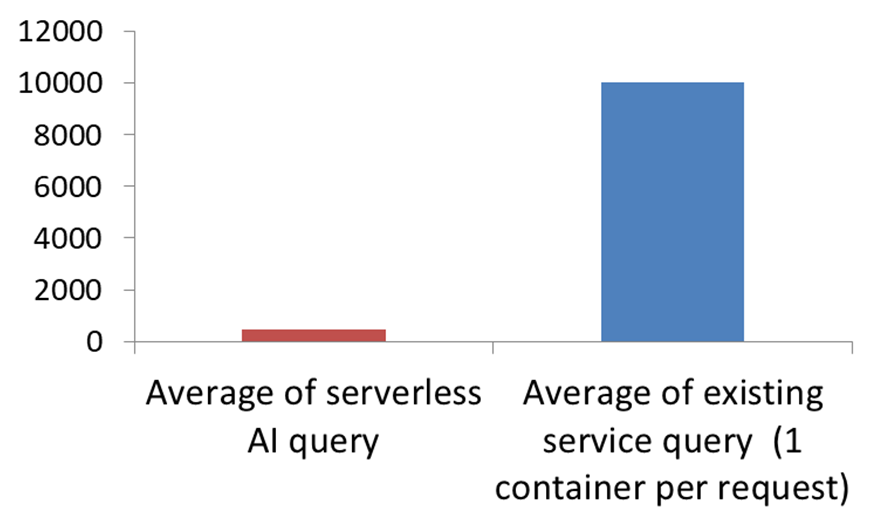}
\caption{Comparison of request delay for 1 request per second case in the FaaS vs AI service implementation of  \cite{Kousiouris2021self}}
\label{fig:octave_measurement}
\end{center}
\end{figure}

\subsection{Example C-Function Orchestartor for the typical Fork-Join Pattern}

The Fork-Join pattern presented in Section III is offered as a subflow and can thus be included in any Node-RED flow. However in order to fully exploit the functionality of the environment, the flow can be executed in a function mode, thus enabling higher scalability of the orchestration mechanism.

In order to enable this, the subflow needs to be wrapped around the Openwhisk Skeleton flow. Arguments are passed by the Openwhisk runtime as the body of the POST /run method. The needed arguments are two, initially the name of the inner Openwhisk action to invoke (the function responsible for processing each input chunk) and the initial array of input data that are split and sent for processing in chunks. The resulting flow appears in Fig. \ref{fig:splitjoinorch} . The latter includes also a couple of testing flows, that can be used while the developer implements and tests the flow inside the Node-RED editor environment. These can prove very useful for debugging and fixing a number of Javascript or data passing related errors without having to deploy to the actual Openwhisk environment. Indicatively, during the tests for the creation of this flow, approximately 70-80\% of the errors were fixed without the need to deploy to the final FaaS platform.

\begin{figure}[htbp]
\includegraphics[width=\columnwidth]{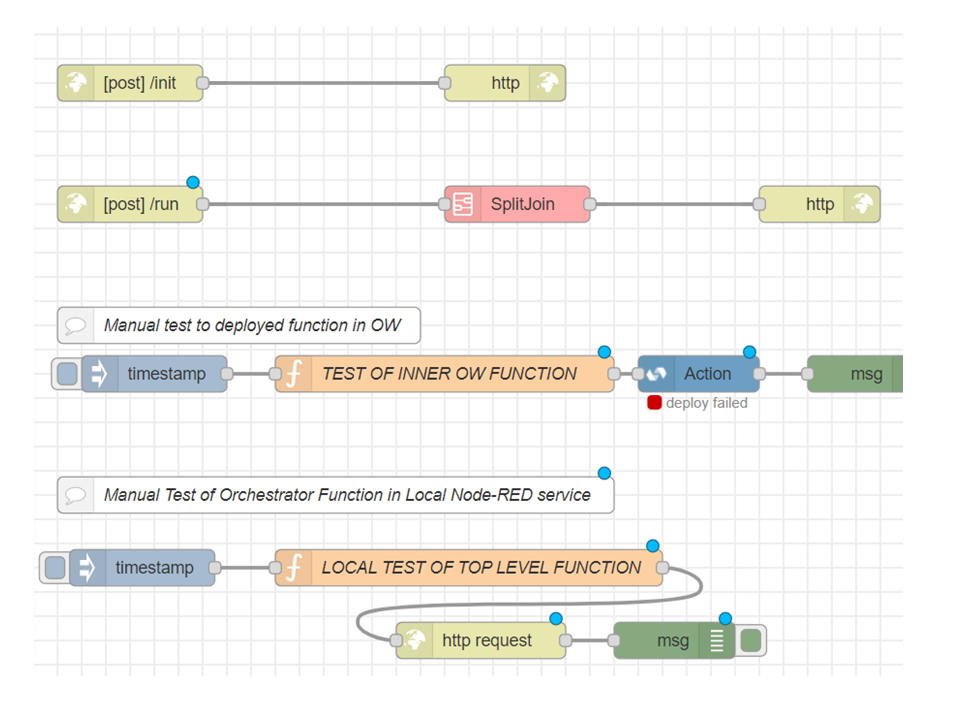}
\caption{Orchestrator Function of the Split Join Pattern including local testing}
\label{fig:splitjoinorch}
\end{figure}

\subsection{Example D- Edge ETL service for Smart Agriculture}

As mentioned in the previous sections, the goal of the platform is to combine function and service implementations, for creating applications that exploit the best of the two worlds. A designed Node-RED flow can be deployed as a function or a service, depending on the requirements. For the case of data collection, especially in resource constrained environments such as Smart Agriculture greenhouses, the edge device (typically a device such as a Raspberry Pi) may be too constrained to run a FaaS platform, even a light version of it such as OWL\cite{OWlight}. For this reason, the inclusion of flows as services in a Node-RED server may be the best way to utilize the resource. The usage of Node-RED  is specifically important since the specific tool originates from the IoT domain and includes an abundance of interacting nodes  with many respective sensor protocols.

However, in these cases, other problems may exist (e.g. volatility of networking conditions), which results in frequent network failures and missing values. Completeness of these data are key in order to run agronomic simulations and calculate the necessary management of the plants. To this end, a  flexible Edge Extract-Transform-Load (ETL) pattern has been  developed for obtaining the baseline sensor data and sending them to a centralized cloud storage and analysis service. 

The design and rationale of the pattern appears in Fig.\ref{fig:edgeservice} . The pattern receives the data item through an incoming message (the input layer can use whatever available Node-RED node to interact with the main IoT system) and tries initially to send the data item to the central service for a limited and configurable number of consecutive times (e.g. 5). If these attempts fail, the data item is stored in a local sqlite node. Periodically, a cron job can also be set to try to resend all failed items up to this point. Other optimizations can also be applied, e.g. testing the network connection before attempting to resend all the contents of the local mini DB.

\begin{figure}[htbp]
\includegraphics[width=\columnwidth]{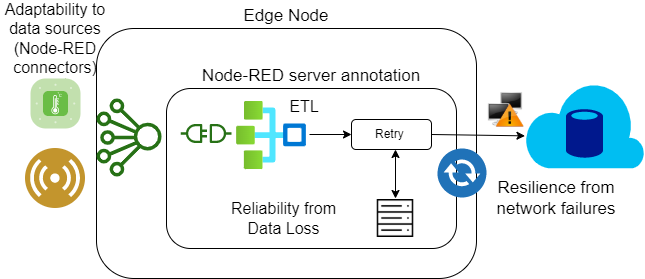}
\caption{Edge ETL Service Conceptual Design}
\label{fig:edgeservice}
\end{figure}

The according Node-RED subflow implementation appears in Fig. \ref{fig:edgeflow}, including approximately 23 inner javascript functions. The usage of the pattern in an arbitrary flow appears in Fig. \ref{fig:edgeexec}, assuming an HTTP endpoint for getting the data, as well as using a semantic node (Executor Mode node) that indicates that this flow needs to be deployed as a service.  

\begin{figure}[htbp]
\includegraphics[width=\columnwidth]{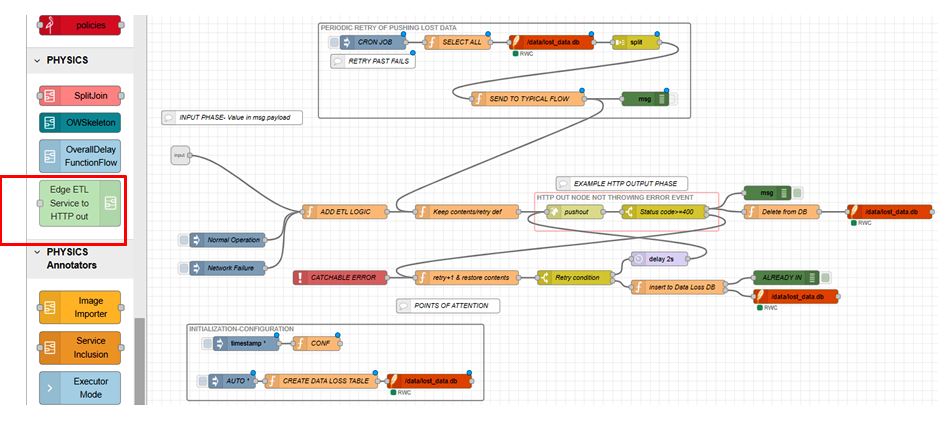}
\caption{Edge ETL Service Subflow}
\label{fig:edgeflow}
\end{figure}

\begin{figure}[htbp]
\includegraphics[width=\columnwidth]{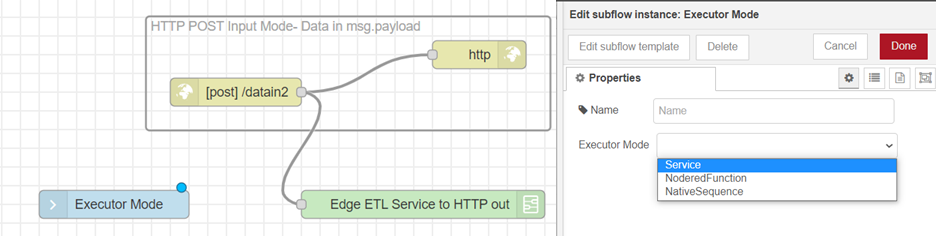}
\caption{Usage of Edge ETL subflow and execution mode annotation}
\label{fig:edgeexec}
\end{figure}

For testing the pattern, two aspects are important. Initially how the pattern scales in terms of accumulated messages during failures and finally whether the application of the pattern results in no lost data. For testing this, the cloud storage service target URL was set to change every 5 minutes to an existing endpoint and every 3 minutes to a non-existent one. The difference in the intervals thus generates a condition simulating the network outage. 

A client generated 2 calls per second for creating new data items. The periodic cron job for past failed data submission was set to 2 minutes. During the experiment, log files documented the calls based on a unique id at both the edge and cloud sides. Thus at the end of the experiment one can join the log files to detect missing values. The experiment was left to execute for ~2 hours. The evolution of the db size appears in Fig. \ref{fig:edgeetl}, fluctuating as expected from the difference in the simulated outage, and reaching a large number of points (\~1400). This generates a burst at the edge side when the periodic cron job is activated, which however is manageable and results in cleaning the edge database in each  cycle. 

\begin{figure}[htbp]
\begin{center}
\includegraphics[scale=0.25]{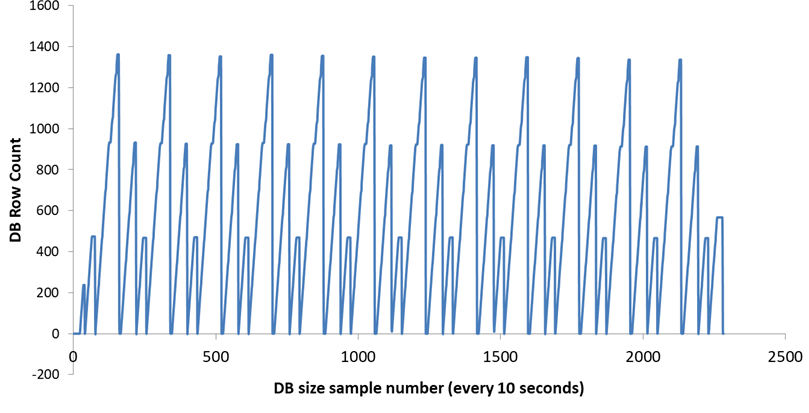}
\caption{Evolution of Failed Data DB size during experiment}
\label{fig:edgeetl}
\end{center}
\end{figure}

From the analysis of log files it was concluded that no single data item was lost from the 18500 samples. Indicatively, the Smart Agriculture Use Case in the context of the PHYSICS project\footnote{https://physics-faas.eu/} reports that the pre-PHYSICS implementation loses approximately 50\% of the available data items. This results to high inaccuracies in the agronomic models application.

However even in this case one may need to execute this pattern as a function. For adapting it, the main modifications are to change from the usage of the local mini DB to a function-external edge-local storing mechanism as well as split the two processes (individual data item processing and periodic resubmission) to two functions. The second can be directly applied based on triggers and rules of the FaaS platform \cite{OWscheduled}.

\section{Conclusions}

Concluding, the current implementations in FaaS exhibit a somewhat static and limited view on function workflows, hindering the manageability of functions as well as making  application porting into the paradigm more complex. The PHYSICS Design Environment builds upon Node-RED capabilities and offers a set of enablers on top for leveraging it as a generic function and workflow creator. 

The environment includes dynamic and adaptable DevOps processes for generating the diverse deployable artefacts (native functions or complete collections executed on a newly created Node-RED runtime). Function and service modes are foreseen, for adapting to requirements, while versatile annotations can be used for dictating developer options to the underlying management stacks. Flows are built on top of provided, reusable and configurable pattern subflows. These are easily dragged and dropped in new flow implementations, with minimal needed knowledge on the internals of the pattern, reducing the development time and learning curve.

Basing the flow creation and orchestration on the Node-RED runtime alleviates from the complex, not scalable (in terms of complexity) syntax of typical workflow mechanisms in current FaaS platforms. It also enables the usage of more complex workflow primitives compared to plain Openwhisk sequences. Different implemented examples have been created and tested in the environment for demonstrating the baseline abilities.

For the future, the aim is to extend the implemented patterns and annotations. Furthermore, translation from the Node-RED flow description to more underlying FaaS platform specifications can be implemented as plugins to the DevOps processes or new interface nodes. Finally, link of patterns to runtime modelling mechanisms for identifying key parameters (e.g. Split-Join size of split) would lead to optimized runtime management.

\section*{Acknowledgment}
The research presented has received funding from the European Union's Project H2020 PHYSICS (GA 101017047).

\bibliographystyle{ieeetr}
\bibliography{bib}
\end{document}